\documentclass[aps,twocolumn,nopacs,superscriptaddress,preprintnumbers,prb]{revtex4-1}
\usepackage{graphicx}
\usepackage{bm}
\usepackage{bbold}
\usepackage{amsmath}

\usepackage{epsfig}
\usepackage{physics}
\usepackage{epstopdf}
\usepackage[colorlinks=true,
            linkcolor = blue,
            urlcolor  = blue,
            citecolor = blue,
            anchorcolor = blue]{hyperref}
\usepackage{url}
\usepackage{color}
\usepackage{multirow}
\usepackage{comment}
\usepackage{mathrsfs}
\begin{document}

\title{Discovery of superconductivity in technetium-borides at moderate pressures}
\author{Xiangru Tao}
    \thanks{These authors contributed equally to this work}
\affiliation{MOE Key Laboratory for Non-equilibrium Synthesis and Modulation of Condensed Matter, Shaanxi Province Key Laboratory of Advanced Functional Materials and Mesoscopic Physics, School of Physics, Xi'an Jiaotong University, 710049, Xi'an, Shaanxi, P.R.China}

\author{Aiqin Yang}
    \thanks{These authors contributed equally to this work}
\affiliation{MOE Key Laboratory for Non-equilibrium Synthesis and Modulation of Condensed Matter, Shaanxi Province Key Laboratory of Advanced Functional Materials and Mesoscopic Physics, School of Physics, Xi'an Jiaotong University, 710049, Xi'an, Shaanxi, P.R.China}

\author{Yundi Quan}
\affiliation{MOE Key Laboratory for Non-equilibrium Synthesis and Modulation of Condensed Matter, Shaanxi Province Key Laboratory of Advanced Functional Materials and Mesoscopic Physics, School of Physics, Xi'an Jiaotong University, 710049, Xi'an, Shaanxi, P.R.China}

\author{Biao Wan}
\affiliation{Key Laboratory of Material Physics of Ministry of Education, School of Physics and Microelectronics, Zhengzhou University, Zhengzhou 450052, Henan, P.R.China}

\author{Shuxiang Yang}
\email{yang\_shuxiang@zhejianglab.com\\}
\affiliation{Zhejiang laboratory, Hangzhou, Zhejiang, P.R.China}

\author{Peng Zhang}
\email{zpantz@mail.xjtu.edu.cn\\}
\affiliation{MOE Key Laboratory for Non-equilibrium Synthesis and Modulation of Condensed Matter, Shaanxi Province Key Laboratory of Advanced Functional Materials and Mesoscopic Physics, School of Physics, Xi'an Jiaotong University, 710049, Xi'an, Shaanxi, P.R.China}

\date{\today}

\begin{abstract}
Advances in theoretical calculations boosted the searches for high temperature superconductors, such as sulfur hydrides and rare-earth polyhydrides. However, the required extremely high pressures for stabilizing these superconductors handicapped further implementations. Based upon thorough structural searches, we identified series of unprecedented superconducting technetium-borides at moderate pressures, including TcB (P6$_3$/mmc) with superconducting transition temperature $T_{\text{c}}$ = 20.2 K at ambient pressure and TcB$_2$ (P6/mmm) with $T_{\text{c}}$ = 23.1 K at 20 GPa. Superconductivity in these technetium-borides mainly originates from the coupling between the low frequency vibrations of technetium-atoms and the dominant technetium-4d electrons at the Fermi level. Our works therefore present a fresh group in the family of superconducting borides, whose diversified crystal structures suggest rich possibilities in discovery of other superconducting transition-metal-borides. 
\end{abstract}

\maketitle
\section{Introduction}
The discovery of superconductivity in mercury \cite{onnes1911} motivated a centurial race for superconductors of higher temperatures. 
%Wigner and Huntington initiatively suggest that hydrogen will turn in to metal at high pressure, \cite{RN2} then later in 1968 Ashcroft further suggest that the metallic hydrogen could be a superconductor with high transition temperature. \cite{RN3} Unfortunately, the searching for metallic hydrogen is rather twisty. \cite{Hydrogen.nphys2007,Hydrogen.PRB2012,Hydrogen.PRL2015,Hydrogen.nature2016,Hydrogen.PRL2018} There is no consensus on the pressure conditions at which hydrogen will enter the metallic phase. \cite{RN4,RN5,RN6} 
%In 2004, Ashcroft proposed that hydrogen atoms in compounds could be pre-compressed by other atoms, thus lower the required physical pressure for superconductivity. \cite{RN7} %
Owing to the progress in theoretical calculations, \cite{RN8,RN9,RN10,RN11} numerous high-temperature superconducting hydrides have been discovered in the past decade, including H$_3$S ($T_{\text{c}}$ $\approx$ 191 - 204 K at 200 GPa) \cite{Duan.SciRept2014,Drozdov.Nature2015} and LaH$_{10}$ ($T_{\text{c}}$ $\approx$ 274 - 286 K at 210 GPa) of record high superconducting transition temperature. \cite{Liu.PNAS2017,Geballe.Angew2018,Somayazulu.PRL2019, Drozdov.Nature2019} %Today, the discovered hydrogen rich superconductors include binary hydrides like sulfur hydride \cite{Duan.SciRept2014,Drozdov.Nature2015}, rare earth hydrides \cite{Peng.PRL2017}, alkali metal hydrides \cite{RN19}, alkali earth metal hydrides \cite{RN20}, transition-metal hydrides etc. \cite{RN21} and ternary hydrides \cite{RN22}. 
However, stable presence of these superconducting hydrides requests very high pressures, which largely limits their potential implementations. %Although recent paper by Dasenbrock-Gammon {\it et al.} \cite{Dias.Nature2023} claimed the discovery of $T_{\text{c}}$ = 294 K at 1 GPa in lutetium-nitrogen-hydrogen, most subsequent experimental and theoretical works refute their conclusions. \cite{Shan.CPL2023, Ming.arxiv2023, Xing.arxiv2023, Cai.MRE2023, Salke.arxiv2023, DPeng.MRE2023, Tao.SciBull2023, HuoZihao.MRE2023} 

Among all BCS-superconductors, borides represent a unique category with superconductivity at relatively low pressures. MgB$_2$ has the highest superconducting transition temperature, $T_{\text{c}}$=39 K, among all BCS-type superconductors at ambient pressure. \cite{Nagamatsu.Nature2001} Up to today, discovered bulk superconducting borides of the same stoichiometry as MgB$_2$ include CaB$_{2}$ ($T_{\text{c}}$ $\sim$ 50 K \cite{Choi.PhysRevB.80.064503} or 9.4 - 28.6 K \cite{Yu.PRB2022} at ambient pressure, theory), NbB$_2$ ($T_{\text{c}}$ $\sim$ 9.2 K at ambient pressure, experiment \cite{Schirber.PRB1992, Yamamoto.PCS2002, Takeya.PCS2004}), OsB$_{2}$ ($T_{\text{c}}$=2.1 K at ambient pressure, experiment \cite{Singh.PhysRevB.76.214510}), RuB$_{2}$ ($T_{\text{c}}$=1.6 K at ambient pressure, experiment \cite{Singh.PhysRevB.76.214510}), ScB$_2$ ($T_{\text{c}}$=1.5 K at ambient pressure, experiment \cite{Samsonov.1980}), 
 WB$_2$ (maximum $T_{\text{c}}$=15 K at 100 GPa, experiment \cite{Pei.SCPMA2022}), 
 ZrB$_2$ ($T_{\text{c}}$=5.5 K at ambient pressure, experiment \cite{Leyarovska.1979}). Superconducting borides of other stoichiometry include X$_{7}$B$_{3}$ (X=Re and Ru with $T_{\text{c}}$=3.3 and 2.58 K respectively at ambient pressure, experiment \cite{Buzea.SST2001, Kawano.JPSJ2003}), Re$_{3}$B ($T_{\text{c}}$=4.8 K at ambient pressure, experiment \cite{Kawano.JPSJ2003}), X$_{2}$B (X=Mo, Re, Ta and W with $T_{\text{c}}$=5.07, 2.8, 3.12 and 3.22 K respectively at ambient pressure, experiment \cite{Buzea.SST2001}), XB (X=Hf, Nb, Mo, Ta and Zr with $T_{\text{c}}$=3.1, 8.25, 0.5, 4.0 and 2.8-3.4 K respectively at ambient pressure, experiment \cite{Buzea.SST2001}), FeB$_{4}$ ($T_{\text{c}}$=2.9 K at ambient pressure, theory and experiment \cite{Kolmogorov.PRL2012, Gou.PRL2013}), XB$_{5}$ (X=Na, K, Rb, Ca, Sr, Ba, Sc and Y with $T_{\text{c}}$=17.5, 14.7, 18.6, 6.6, 6.8, 16.3, 14.2 and 12.3 K respectively at ambient pressure, theory \cite{Xie.MRE2023}), BeB$_6$ ($T_{\text{c}}$=24 K at 4 GPa, theory \cite{Wu.JPCL2016}), CB$_6$ ($T_{\text{c}}$=12.5 K at ambient pressure, theory \cite{Xia.MTP2017}), MgB$_{6}$ ($T_{\text{c}}$=9.5 K at 32.6 GPa, theory \cite{Duan.DT2019}), XB$_{6}$ (X=Nb, La, Th and Y with $T_{\text{c}}$=3.0, 5.7, 0.74 and 7.1 K respectively at ambient pressure, experiment \cite{Buzea.SST2001}), XB$_{7}$ (X=Li, Na, K, Mg, Ca and Sr with $T_{\text{c}}$=21.56, 18.33, 26.20, 29.31, 7.68 and 12.67 K respectively at ambient pressure, theory \cite{Han.AFM2023}), RbB$_{6}$ and RbB$_{8}$ ($T_{\text{c}}$=7.3 - 11.6 K and 4.8 - 7.5 K at ambient pressure respectively, theory \cite{Zhang.PRR2023}), YB$_{6}$ ($T_{\text{c}}$=7.2 K at ambient pressure, experiment \cite{Lortz.PhysRevB.73.024512}), LaB$_8$ ($T_{\text{c}}$=14 K \cite{Ma.PhysRevB.104.174112} or 20 K \cite{Liang.JMCC2021} at ambient pressure, theory), XB$_{12}$ (X=Nb, La, Th and Y with $T_{\text{c}}$=3.0, 5.7, 0.74 and 7.1 K respectively at ambient pressure, experiment \cite{Buzea.SST2001,Akopov.jacs2019,Teyssier.PhysRevB.75.134503}), or even ternary borides like SrB$_{3}$C$_{3}$ ($T_{\text{c}}$=22 K at 23 GPa, theory and experiment \cite{Zhu.PRR2023}). Especially, the recent experimental discovery of superconducting MoB$_2$ with $T_{\text{c}}$=32 K at 100 GPa ignites further enthusiasm in looking for superconducting transition-metal-borides at ambient pressure or at least relatively lower pressures.\cite{Pei.NSR2023} 
 
 The potential superhardness of metal-borides further merits their values. Among the superconducting metal-borides listed above, several are superhard due to the strong covalent bonding between the metal-boron and the boron-boron atoms in crystals. Superconducting FeB$_4$ has nanoindentation hardness of 62$\pm5$ GPa. \cite{Gou.PRL2013} 
 Other superconducting metal-borides with superhardness includes OsB$_{2}$ ($\geq$ 2000 kg/mm$^2$, \cite{Cumberland.JACS2005}) YB$_{6}$ (Vickers hardness of 37.0 GPa \cite{Ding.JPCL2021}), 
 RbB$_{6}$ (Vickers hardness of 19.7 GPa \cite{Zhang.PRR2023}), 
 RbB$_{8}$ (Vickers hardness of 36.9 GPa \cite{Zhang.PRR2023}), 
 XB$_{7}$ (X=Li, Na, K, Mg, Ca, Sr with Vickers hardness of 12.0, 21.3, 22.5, 5.6, 20.4 and 25.1 GPa \cite{Han.AFM2023}) and ZrB$_{12}$ (Vickers hardness of 40 GPa \cite{Ma.AM2017}).

\begin{figure*}[tbp]
  \centering
  \includegraphics[width=1.0\textwidth]{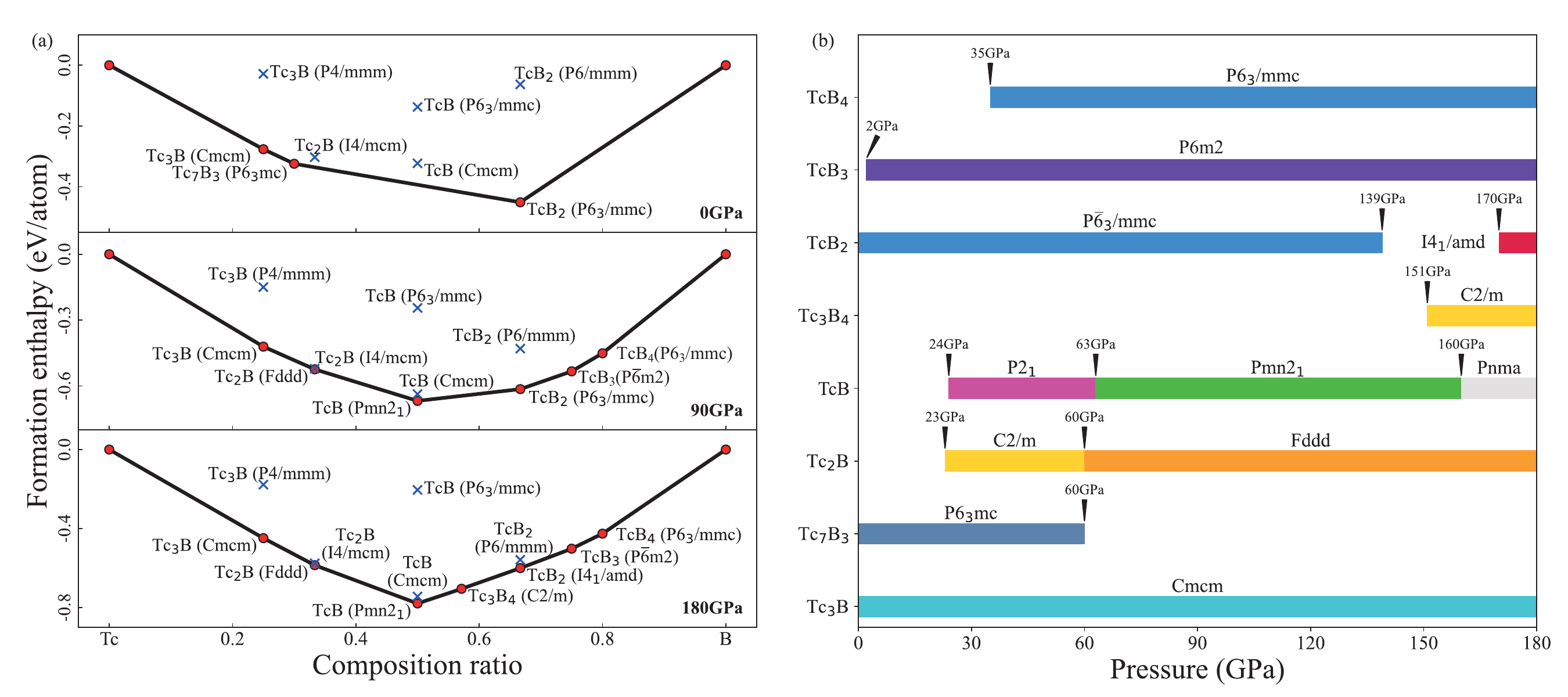}
  \caption{(a) Formation enthalpy of predicted structures in technetium-boron binary system at 0 GPa, 90 GPa and 180 GPa. Thermodynamically stable structures are marked by red-filled dots on convex hull (black solid line); thermodynamically meta-stable structures are marked by blue 'x'. Composition ratio is defined by $N_B/(N_{T_c}+N_B$), where $N_B$ and $N_{T_c}$ represent the number of atoms in the formula unit. (b) Composition-pressure phase diagram of thermodynamically stable structures in technetium-boron binary system.}
  \label{convexhull}
\end{figure*}

After the discovery of $MgB_2$ with $T_c$=39 K, there have been extensive efforts in searching for similar metal-borides superconductors. Unfortunately, the outcomes are discouraging. $T_c$ of most metal-borides are below 10 K as presented above, except the recently discovered isostructure $MoB_2$ with $T_c$=32 K at 100 GPa. The superconducting mechanism of $MoB_2$ is suggested being very different from $MgB_2$. In $MgB_2$ the B-p electrons play dominant role for its superconductivity, \cite{An.PRL.86.4366, Kortus.PRL2001} while in $MoB_2$ its Mo-4d electrons contribute majorly. \cite{Pei.NSR2023} This raises two essential questions, (1) can we found other superconductors of different structures than $MgB_2$, with transition temperatures at least above 10 K and at not very high pressures? (2) whether the superconducting mechanism of $MoB_2$ applies to other superconducting transition metal-borides?

Technetium-borides have been extensively investigated because of their outstanding mechanical properties. \cite{Trzebiatowski.JLCM1964, ARMSTRONG.JLCM1979, Pallas.JPCB2006, Wang.APL2007, Wang.APL2008, Aydin.PRB2009, Li.PBCM2010, Chen.SSC2010, Zhao.CMS2012, Deligoz.SSC2012, Zhong.JPCC2013, Zhang.CMS2013, Ying.MPLB2014, Kuang.CIC2015, VANDERGEEST.2014, Wu.CMS2014, Zhang.CPB2015, Ying.IJMPB2016, Miao.SSC2017, Ying.CMS2018, Wu.APA2023} Three technetium-borides have long been synthesized by experiment at ambient pressure, \cite{Trzebiatowski.JLCM1964} Tc$_3$B (Cmcm) of the orthorhombic structure, Tc$_7$B$_3$ (P6$_3$/mmc) and TcB$_2$ (P6$_3$/mmc, Vickers hardness 38.4 GPa \cite{Wu.APA2023} or 39.4 GPa \cite{Ying.CMS2018}) of the hexagonal structure. Later theoretical calculations also proposed three stoichiometry of TcB, TcB$_3$ and TcB$_4$. \cite{Li.PBCM2010, Wu.CMS2014,Zhang.CPB2015,VANDERGEEST.2014,Miao.SSC2017,Ying.IJMPB2016, Ying.CMS2018} First-principle DFT calculations by Li {\it et al.} \cite{Li.PBCM2010} suggest that hexagonal TcB (P$\Bar{6}$m2) could be energetically stable. Structural searches by Wu {\it et al.} \cite{Wu.CMS2014} found a thermodynamically stable TcB (Cmcm) structure above 8 GPa. Later structrual searches by Zhang {\it et al.} \cite{Zhang.CPB2015} argues that TcB (P$\Bar{3}$m1, Vickers hardness 30.3 GPa) could be energetically more stable than the above two structures. Structural predictions by Van Der Geest {\it et al.} \cite{VANDERGEEST.2014} suggest there are two thermodynamically stable structures, TcB(Pnma) and TcB$_4$ (P6$_3$/mmc), at 30 GPa. First-principle DFT calculations by Miao {\it et al.} \cite{Miao.SSC2017} reported thermodynamically stable TcB$_3$ (P$\Bar{6}$m2, Vickers hardness 29 GPa) structure at above 4 GPa. Structural searches by Ying {\it et al.} \cite{Ying.IJMPB2016, Ying.CMS2018} suggested two structures, TcB$_3$ (P$\Bar{6}$m2, Vickers hardness 30.7 GPa) and TcB$_4$ (P6$_3$/mmc, Vickers hardness 32.4 GPa), are thermodynamically stable at 0 and 100 GPa respectively.

Although technetium is rare in nature, the technetium-based compounds are under investigations by multiple disciplines in the past. Perovskites ATcO$_3$ (A = Ca, Sr, Ba) attracted extensive interests due to their extremely high antiferromagnetic Neel temperatures (750–1200 K). \cite{Avdeev.JACS2011, Franchini.PRB2011, Rodriguez.PRL2011, Mravlje.PRL2012} Recently, techentium hydrides were theoretically predicted and then experimentally synthesized under high pressure. \cite{LiXiaofeng.PCCP2016, ZhouDi.PRB2023}

Previous works on technetium-borides are concentrated on the high hardness and the high incompressibility of technetium-borides, while the explorations for superconductivity are absent. Considering the fact that series of transition-metal-borides have been found superconducting at relatively low pressures, there is no reason to rule out the possible presence of superconductivity in technetium-borides. Besides, current results about the thermaldynamically stable phases of technetium-borides are highly diversified. Therefore, we choose to search the technetium-boron binary system for new superconductors at low or even ambient pressures. A comprehensive phase diagram of all thermodynamically stable technetium-borides up to 180 GPa has been derived. We also found five new superconducting technetium-borides of metastable states, including TcB (P6$_3$/mmc), TcB$_2$ (P6/mmm), Tc$_2$B (I4/mcm), Tc$_3$B (P4/mmm) and TcB (Cmcm), which remain dynamically stable at relatively low pressures. The mechanical properties of these superconducting technetium-borides have been investigated as well.

\section{Methods}

The structure prediction for technetium-boron binary crystals is performed by CALYPSO package\cite{CALYPSO}. The electronic structures and the phonon properties are calculated using QUANTUM-ESPRESSO (QE) package \cite{QE}. The plane-wave kinetic-energy cutoff and the charge density energy cutoff are 100 Ry and 400 Ry, respectively. Optimized norm-conversing pseudopotential with valence electron configurations of Tc-4p$^6$4d$^5$5s$^2$ and B-2s$^2$2p$^1$ and Methfessel-Paxton smearing \cite{RN56} width of 0.02 Ry are used. Since technetium is radioactive without stable isotopes, we adopted the average mass of technetium and boron at 97.907 and 10.811 atomic mass units throughout our calculations. The dynamic matrix and the electron-phonon coupling (EPC) constant $\lambda$ are calculated using the density-functional perturbation theory \cite{DFPT}. Superconducting transition temperature is estimated following the Allen-Dynes modified McMillan equation \cite{Allen.PR1975},

\begin{eqnarray}
T_{\text{c}} = \frac{\omega_{\text{log}}}{1.2} \exp \left [ -\frac{1.04(1+\lambda)}{\lambda - \mu^\ast\left (1+0.62 \lambda 
 \right ) }\right ],
\end{eqnarray}

in which $\lambda$ is the average EPC parameter, $\omega_{\text{log}}$ is the logarithmic average frequency, and the Coulomb pseudopotential \cite{Morel.PR1962} $\mu^{*}$ = 0.12. Mechanical properties including Vickers hardness are estimated following models by Chen {\it et al.} and Tian {\it et al.} \cite{Chen.Intermetallics2011,Tian.MetH2012}. Calculation details are referenced to the supplementary information \cite{SI}.

\section{Results and discussion}
\subsection{Convex hull and phase diagram}

\begin{figure*}[htbp]
  \centering
  \includegraphics[width=0.98\textwidth]{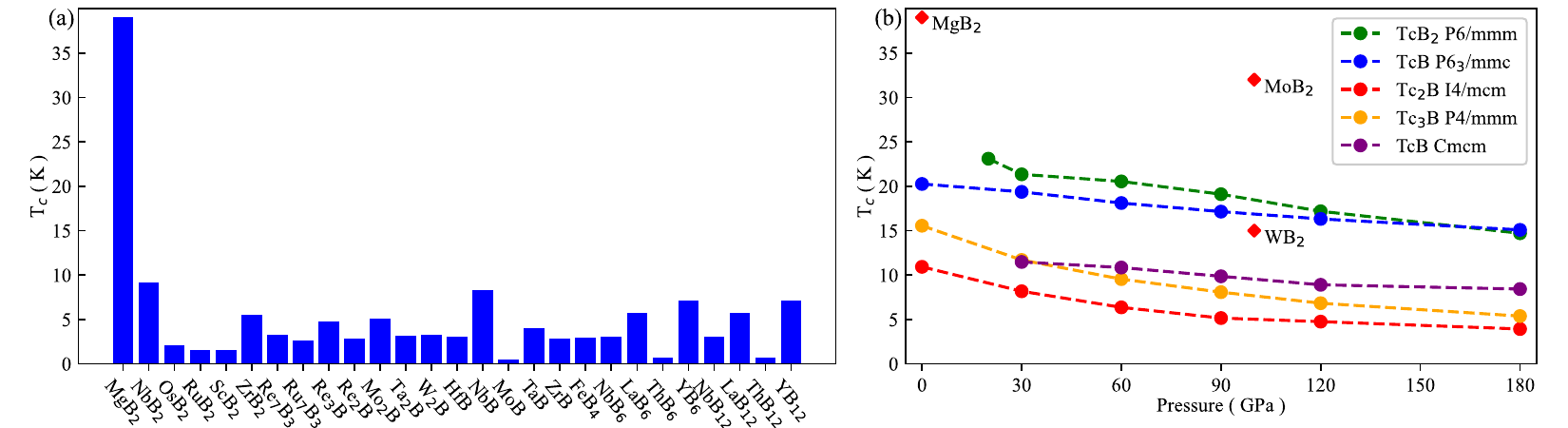}
  \caption{Superconducting transition temperature of TcB$_2$ (P6/mmm), TcB (P6$_3$/mmc), Tc$_2$B (I4/mcm), Tc$_3$B (P4/mmm) and TcB (Cmcm) as a function of pressure.}
  \label{Tc-P}
\end{figure*}

We have done variable-composition and fixed-composition structure searches in the Tc-B system at pressures of 0, 90 and 180 GPa. Thermodynamically stable structures and the derived composition-pressure phase diagram are presented in Fig.~\ref{convexhull}. Three existing technetium-borides at ambient pressure, Tc$_3$B (Cmcm), Tc$_7$B$_3$ (P6$_3$/mmc) and TcB$_2$ (P6$_3$/mmc), have been successfully identified. Tc$_3$B (Cmcm) is thermodynamically stable up to 180 GPa in our study. In contrast, Tc$_7$B$_3$ (P6$_3$/mmc) and TcB$_2$ (P6$_3$/mmc) stop being energetically favorable above 60 GPa and 139 GPa respectively, and a new TcB$_2$ (I4$_1$/amd) thermodynamically stable phase shows up above 170 GPa. Two previously predicted structures, TcB$_3$ (P$\Bar{6}$m2) \cite{Miao.SSC2017,Ying.IJMPB2016} and TcB$_4$ (P6$_3$/mmc) \cite{VANDERGEEST.2014,Ying.CMS2018}, also have been found in our calculations, which are thermodynamically stable above 2 and 35 GPa, respectively. We found TcB (P2$_1$) structure being thermodynamically stable above 24 GPa, then transfers into Pmn2$_1$ structure at 63 GPa, and finally into the previously predicted Pnma structure \cite{VANDERGEEST.2014} at 160 GPa. We also discovered technetium-borides of two new stoichiometries, Tc$_3$B$_4$ and Tc$_2$B. Tc$_3$B$_4$ (C2/m) is thermodynamically stable above 151 GPa. Tc$_2$B (C2/m) structure is thermodynamically stable above 23 GPa, then transfers into Fddd structure at 60 GPa. The crystal structure information of all thermodynamically stable phases are presented in Table IV of supplementary information \cite{SI}. 

\begin{table}
    \caption{Total electronic DOS at Fermi level N(E$_F$), EPC parameter $\lambda$, logarithmic average frequency $\omega_{log}$, and superconducting transition temperature $T_{\text{c}}$ of the superconducting technetium-borides at their lowest dynamically stable pressures.}
    \label{Tc_lowestP}%
    \begin{ruledtabular}
        \begin{tabular}{cccccccc}
        Formula & Space       & P      & N(E$_F$)          & $\lambda$     & $\omega_{log}$    & $T_{\text{c}}$   \\
             &  group      & (GPa)  &(states/eV/f.u.)   &               & (cm$^{-1}$)         & (K)     \\ 
        \hline
        TcB$_2$  & P6/mmm      & 20     &    1.41           &  1.85         &    125.1           & 23.1   \\
        TcB      & P6$_3$/mmc  & 0      &    1.63           &  1.56         &    126.1           & 20.2   \\
        Tc$_2$B  & I4/mcm      & 0      &    1.75           &  0.85         &    165.1           & 10.9   \\
        Tc$_3$B  &  P4/mmm     & 0      &    2.85           &  0.92         &    162.9           & 12.9   \\
        TcB      &  Cmcm       & 30     &    1.05           &  0.96         &    135.5           & 11.5   \\
        \end{tabular}%
    \end{ruledtabular}
\end{table}

\begin{figure*}[ht]
 \includegraphics[width=0.98\textwidth]{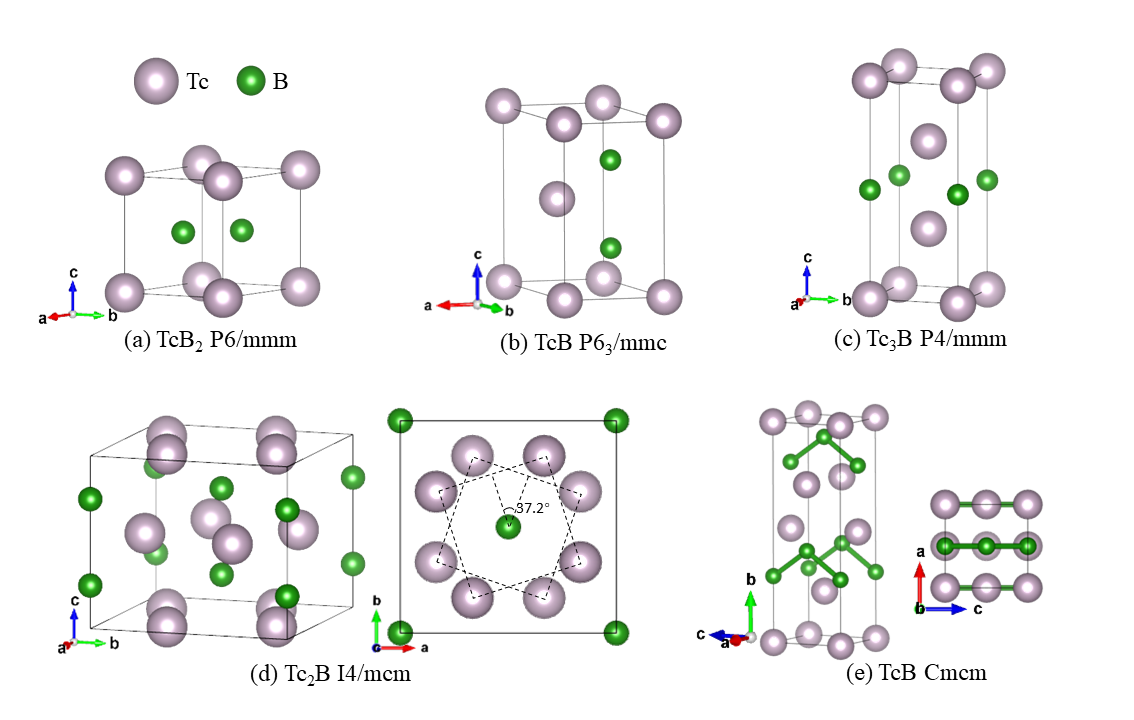} 
  \caption{Crystal structures of superconducting technetium-borides. (a) TcB$_2$ (P6/mmm) (b) TcB (P6$_3$/mmc) (c) Tc$_2$B (I4/mcm) (d) Tc$_3$B (P4/mmm) and (e) TcB (Cmcm). The technetium and the boron atoms are represented by spheres of coral and green colors, respectively.}
  \label{structures}
\end{figure*}

\subsection{Superconductivity of thermodynamically metastable technetium-borides}

We also examined potential superconductors in technetium-borides, including all thermodynamically stable structures and the thermodynamically metastable structures within range of 300 meV above the convex hull. Totally five thermodynamically metastable technetium-borides have been found superconducting at 180 GPa, including TcB$_2$ (P6/mmm, 42 meV/atom above the hull), TcB (P6$_3$/mmc, 255 meV/atom above the hull), Tc$_2$B (I4/mcm, 2 meV/atom above the hull), Tc$_3$B (P4/mmm, 248 meV/atom above the hull) and TcB (Cmcm, 25 meV/atom above the hull). These five superconducting technetium-borides stay dynamically stable at decreased pressures. The minimum dynamical stable pressures of TcB$_2$ (P6/mmm) and TcB (Cmcm) are 20 and 30 GPa respectively, while TcB (P6$_3$/mmc), Tc$_2$B (I4/mcm) and Tc$_3$B (P4/mmm) are dynamically stable even at ambient pressure. 

The superconducting transition temperatures of all five technetium-borides increase at decreased pressure in Fig. ~\ref{Tc-P}. We also summarize the superconducting transition temperature of the five technetium-borides at their lowest dynamically stable pressures, together with their total electronic DOS at the Fermi level N(E$_F$), the EPC parameter $\lambda$ and the logarithmic average frequency $\omega_{log}$ in Table ~\ref{Tc_lowestP}. TcB$_2$ (P6/mmm) has the highest superconducting transition temperature of 23.1 K at 20 GPa, which comes from its largest EPC parameter $\lambda$=1.85. 
In contrast, TcB (Cmcm) has much lower superconducting transition temperature of 11.5 K at 30 GPa due to its small EPC parameter $\lambda$=0.96. Superconducting transition temperature of TcB (P6$_3$/mmc), Tc$_2$B (I4/mcm) and Tc$_3$B (P4/mmm) at 0 GPa are 20.2, 10.9, and 12.9 K, respectively. Although the EPC parameters $\lambda$ of these metastable technetium-borides are not small, their logarithmic average frequency $\omega_{log}$ are rather low at maximumly 165.1 cm$^{-1}$, which limited their superconducting transition temperature. 
This is in sharp contrast with MgB$_2$, which has smaller $\lambda$ = 0.87 but much larger $\omega_{log}$ = 504 cm$^{-1}$ and the highest BCS-type superconducting transition temperature of 39 K at ambient pressure. \cite{Kong.PRB2001}

The thermodynamically metastable nature of the discovered superconducting technetium-borides doesn't necessarily exclude their experimental synthesis. Metastable materials have long been synthesized and implemented, \cite{Aykol.sciadv2018} typically like fullerene C$_{60}$. As to superconductor, DFT calculations predict NdH$_9$ (P6$_3$/mmc) is 35 meV/atom above the convex hull at 150 GPa, yet being successfully synthesized with $T_{\text{c}}$ $\approx$ 4.5 K. \cite{DZhou.jacs2020} Especially, several metastable borides have been predicted superconducting in recent structural searches. Xia {\it et al.} discovered thermodynamically metastable CB$_6$ with superconducting transition temperature of 12.5 K at ambient pressure. \cite{Xia.MTP2017} Zhang {\it et al.} also found thermodynamically metastable RbB$_6$ (Pm-3m) and RbB$_8$ (Immm) with superconducting transition temperatures of 7.3–11.6 and 4.8–7.5 K at ambient pressure, respectively. \cite{Zhang.PRR2023} These works further validate the importance and necessity of our discoveries of superconducting technetium-borides.

\begin{figure*}[!htbp]
 \includegraphics[width=0.95\textwidth]{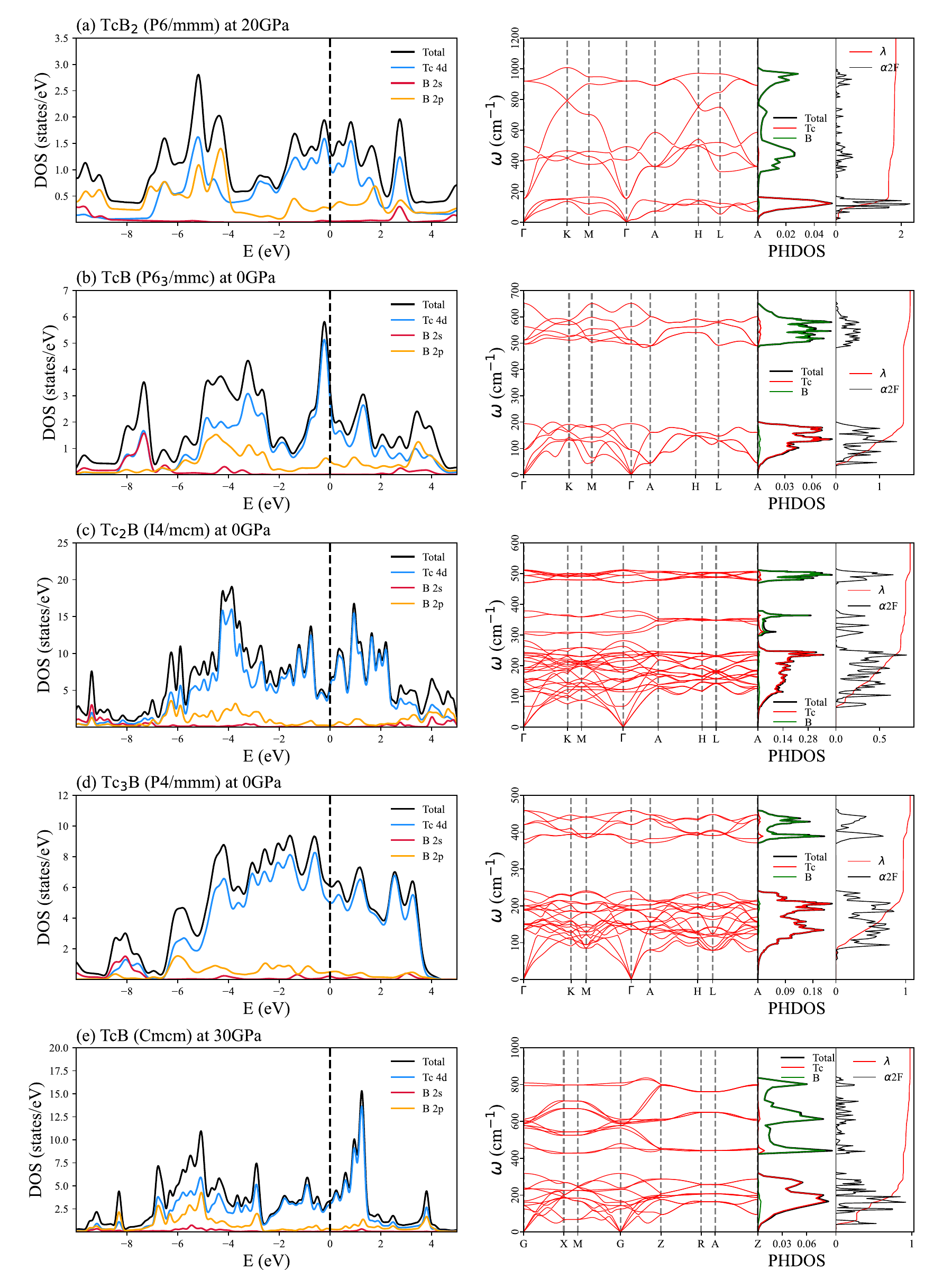} 
  \caption{Total and partial electronic DOS, phonon dispersion relation, phonon density of states (PHDOS), Eliashberg functional $\alpha^2F(\omega)$ and integrated EPC parameter $\lambda(\omega)$ of superconducting technetium-borides at their lowest dynamically stable pressures. From top to bottom, (a) TcB$_2$ (P6/mmm, 20 GPa), (b) TcB (P6$_3$/mmc, 0 GPa), (c) Tc$_2$B (I4/mcm, 0 GPa), (d) Tc$_3$B (P4/mmm, 0 GPa) and (e) TcB (Cmcm, 30 GPa).}
  \label{el-ph}
\end{figure*}

\subsection{Crystal structures}

The crystal structures of five superconducting technetium-borides in our study are presented in Fig.\ref{structures}. TcB$_2$ (P6/mmm) shares exactly the same crystal structure with MgB$_2$ and MoB$_2$. TcB (P6$_3$/mmc) has TiAs-type structure, in which the rhombus Tc-layers are AB-stacking along the c-axis and the rhombus B-layers are sandwiched between the neighbouring Tc-layers. Tc$_3$B (P4/mmm) has square Tc-layers stacking in ABB-pattern along the c-axis, and the square B-layers locate between the two Tc-layers of BB-pattern. Tc$_2$B (I4/mcm) consists of square Tc-layers AB-stacking along the c-axis, where the neighbouring Tc-layers are twisted by 37.2 degrees. The B-layers in Tc$_2$B (I4/mcm) are sandwiched between the neighbouring Tc-layers as well. In TcB (Cmcm), the square Tc-layers stack in ABCD-pattern along the b-axis, and the B-atoms form zig-zag chains along the c-axis between the AB and CD Tc-layers. The angle of the zig-zag chain of B-atoms is around 108.8 degree. Crystal structure information of the superconducting technetium-borides are referenced to Table III in the supplementary information. \cite{SI} 

\subsection{Electronic structures}

The electronic DOS of superconducting technetium-borides at their lowest stabilizing pressure are presented in the left columns of Fig.\ref{el-ph}. The electronic DOS of all technetium-borides share certain features. The total DOS at the Fermi level are dominated by the states of Tc-4d bands. Although the B-2p DOS have considerable weight away the Fermi level, its contribution is minor at the Fermi level, if not being zero. The B-2s DOS almost vanish around the Fermi level, which makes B-2s bands almost irrelevant for electronic conduction. Our DOS results of technetium-borides are in close resemblance to another transition-metal-boride MoB$_2$ \cite{Pei.NSR2023}, while in obvious contrast with the alkali-earth-metal boride MgB$_2$ \cite{An.PRL.86.4366, Kortus.PRL2001} or alkali-metal boride RbB$_6$. \cite{Zhang.PRR2023} In either MgB$_2$ \cite{An.PRL.86.4366, Kortus.PRL2001} or RbB$_6$ \cite{Zhang.PRR2023}, the major DOS at the Fermi level are contributed by the B-p bands. But in either the superconducting technetium-borides of our study or MoB$_2$, \cite{Pei.NSR2023} the 4d electronic states play dominant roles around the Fermi level. 

The electronic band structure and the Fermi surface of superconducting technetium-borides are presented in Fig. S1 of supplementary information \cite{SI}. For all five superconducting technetium-borides, either their band structure or their Fermi surface show obvious electronic dispersion in three dimensions. Typically for example TcB$_2$, which shares the same cyrstal structure as MgB$_2$ and MoB$_2$, has three dimensional Fermi surface like MoB$_2$ \cite{Pei.NSR2023, Quan.PRB2021} while being distinct from the quasi-two-dimensional Fermi surface of MgB$_2$. \cite{Kortus.PRL2001, Choi.Nature2002} 

\subsection{Dynamical stability and electron-phonon coupling}

The phonon spectrum, the PHDOS, the Eliashberg functional $\alpha^2F(\omega)$ and the corresponding integrated EPC constant $\lambda$ of superconducting technetium-borides at their lowest dynamically stable pressures are presented in the right columns of Fig.\ref{el-ph}. There is no sign of imaginary frequency in the phonon spectrum of all five superconducting technetium-borides, which proves the dynamical stability of these structures at the corresponding pressures. The distribution of the PHDOS and the Eliashberg spectral functional $\alpha^2F(\omega)$ of superconducting technetium-borides show clear separation between the low frequency phonon modes of the heavier Tc-atoms and the high frequency phonon modes of the lighter B-atoms. This enables us to separate the integrated EPC constant $\lambda$ into two parts, the EPC from Tc-atoms $\lambda_{T_c}$, and the EPC from B-atoms $\lambda_{B}$. The ratio of EPC from oscillation of Tc-atoms relative to the total EPC, $\lambda_{T_c}$/$\lambda$, are 0.883, 0.910, 0.873, 0.907 and 0.911 for TcB$_2$ (P6/mmm), TcB (P6$_3$/mmc), Tc$_2$B (I4/mcm), Tc$_3$B (P4/mmm) and TcB (Cmcm), respectively. It indicates superconductivity in these five technetium-borides mainly originates from the coupling between the Tc-4d electrons and the low frequency phonon modes of Tc-atoms. At least three isotopes of technetium have reasonably long half lives (Tc-97, Tc-98 and Tc-99 at 4.2$\times 10^{6}$, 6.6$\times 10^{6}$ and 2.13$\times 10^{5}$ years, respectively). Therefore, we suggest experiments on the isotope effects of technetium to examine our prediction.

The superconducting mechanism of our predicted technetium-borides is similar to that in transition-metal-borides MoB$_2$, whose superconductivity mainly originates from the coupling between the Mo-4d electrons and the low frequency Mo-phonon modes. \cite{Pei.NSR2023} However, the superconducting scenarios in alkali-earth-metal boride MgB$_2$ \cite{An.PRL.86.4366, Kortus.PRL2001} and alkali-metal boride RbB$_6$ \cite{Zhang.PRR2023} are very different in that the couplings between B-2p electrons and the high frequency B-phonon modes play dominant roles. 

The presence of BCS superconductivity or not in transition metal borides such as Mn-B, Mo-B, Tc-B, Re-B and Ru-B are likely being related to the magnitude of correlations of d-electrons in the transition metal. In MnB$_2$, Mn-3d electrons are localized due to their strong correlations, which leads to magnetism and the suppression of BCS-superconductivity. In contrast, the superconductivity in MoB$_2$, technetium-borides, Ru$_7$B$_3$ and Re$_7$B$_3$ are preserved where the correlations of d-electrons in Mo, Tc, Ru and Re are relatively smaller.

Observations in phonon spectrum, PHDOS and EPC of superconducting technetium-borides are consistent with their relatively smaller logarithmic average frequency $\omega_{log}$ as listed in Table ~\ref{Tc_lowestP}, since Tc-atoms are much heavier than B-atoms. The enhanced $\lambda$ plus small $\omega_{log}$ characters of TcB$_2$ have also been seen in iso-structural superconductor TlBi$_2$ of heavy atomic mass, \cite{Aiqin.PRB2023} with $\lambda$=1.4, $\omega_{log}$=37 $cm^{-1}$ and rather low $T_{\text{c}}$=5.5 K. Recent work suggests introduction of hydrogen atoms into non-superconducting transition-metal boride Ti$_2$B$_2$ will result in superconducting Ti$_2$B$_2$H$_4$ ($T_{\text{c}}$=48.6 K at ambient pressure), through expansion of the frequency range of phonon spectrum and consequently enlarged electron-phonon coupling \cite{Han.MTP2023}. Similar hydrogenation probably helps in elevating the superconducting transition temperatures of technetium-borides by enlarging $\omega_{log}$.  

Another interesting observation on the discovered superconducting technetium-borides is that the Fermi levels of TcB$_2$ and TcB fall closely above the very peak positions of their DOS, as shown in Fig. \ref{el-ph}. Since the EPC in technetium-borides is controlled by the coupling between the Tc-4d electrons and the oscillation of Tc-atoms, slight hole-doping could lower the Fermi level, thus enhances the effective number of electrons participating into the superconducting pairing and therefore enlarges the superconducting transition temperatures.

\begin{table}
    \caption{Vickers hardness of superconducting technetium-borides at their lowest dynamically stable pressures. }
    \label{tab:elastic}%
    \begin{ruledtabular}
        \begin{tabular}{cccccccccccccc}
        Formula & Space Group & P      &  {H$_{v,Chen}$} \cite{Chen.Intermetallics2011} &  {H$_{v,Tian}$} \cite{Tian.MetH2012}&\\
              &             & (GPa)  &     (GPa)  &  (GPa)    &\\
        \hline
        TcB$_2$   & P6/mmm      &  20    &  10.6      & 12.0        \\
        TcB       & P6$_3$/mmc  &  0     &  2.8      &  4.8        \\   
        Tc$_2$B   & I4/mmm      &  0     & 11.8      & 13.0        \\
        Tc$_3$B   & P4/mmm      &  0     & 10.0      & 11.3        \\
        TcB       & Cmcm        & 30     & 12.2      & 13.7        \\
        \end{tabular}%
    \end{ruledtabular}
\end{table}

% \begin{table}[tbp]
%   \centering
%   \caption{Vickers hardness of superconducting technetium-borides at their lowest dynamically stable pressures. }
%     \begin{tabular}{cccccccccccccc}
%       \hline 
%     Compound & Space Group & P      &  {H$_{v,Chen}$} \cite{Chen.Intermetallics2011} &  {H$_{v,Tian}$} \cite{Tian.MetH2012}&\\
%               &             & (GPa)  &     (GPa)  &  (GPa)    &\\
%     \hline
%     TcB$_2$   & P6/mmm      &  20    &  10.6      & 12.0        \\
%     TcB       & P6$_3$/mmc  &  0     &  2.8      &  4.8        \\   
%     Tc$_2$B   & I4/mmm      &  0     & 11.8      & 13.0        \\
%     Tc$_3$B   & P4/mmm      &  0     & 10.0      & 11.3        \\
%     TcB       & Cmcm        & 30     & 12.2      & 13.7        \\
%     \hline
%     \end{tabular}%
%   \label{tab:elastic}%
% \end{table}%

\subsection{Hardness}

We also calculated the Vickers hardness of discovered superconducting technetium-borides as presented in Table ~\ref{tab:elastic}, including . At ambient pressure, TcB (P6$_3$/mmc), Tc$_2$B (I4/mcm) and Tc$_3$B (P4/mmm) have Vickers hardness values of 2.8-4.8, 11.8-13.0 and 10.0-11.3 GPa, respectively. The Vickers hardness of TcB$_2$ (P6/mmm) and TcB (Cmcm) are 9.8-11.3 GPa and 12.2-13.7 GPa at pressures of 20 and 30 GPa, respectively. The superconducting technetium-borides in our study have relatively lower hardness values than previously stated superconducting borides of superhardness, for example RbB$_6$ (Pm-3m, Vickers hardness of 19.7 GPa at the ambient pressure) and RbB$_8$ (Immm, Vickers hardness of 36.9 GPa at the ambient pressure). \cite{Zhang.PRR2023} Other mechanical parameters including elastic constants C$_{ij}$, bulk modulus B, and shear modulus G at their lowest dynamically stable pressures are also calculated. Mechanical stability criteria \cite{Mouhat.PRB2014} related to the elastic constants of these superconducting technetium-borides are fulfilled as presented in supplementary information \cite{SI}.

% This could originates from the small valence-electron density of the superconducting technetium-borides. The electronic DOS of RbB$_6$ (Pm-3m) and RbB$_8$ (Immm) at the Fermi level N(E$_F$) are 24.2 and 10.4 (states/eV/f.u.) respectively. In contrast, for TcB$_2$ (P6/mmm) and Tc$_3$B (P4/mmm), which have the highest Vickers hardness values in our study, their electronic DOS are 1.43 and 2.85 (states/eV/f.u.) respectively. 

\section{Conclusion}

In summary, we have conducted thorough structural searches in the technetium-boron binary system. An updated composition-pressure phase diagram of technetium-borides up to 180 GPa have been derived, including two new stoichiometries as Tc$_3$B$_4$ and Tc$_2$B. More importantly, we also found five unprecedented superconducting technetium-borides which remain dynamically stable at moderate or even ambient pressures. Among these thermodynamically metastable superconducting technetium-borides, TcB$_2$ (P6/mmm) has the highest superconducting transition temperature of 23.1 K at 20 GPa, and TcB (P6$_3$/mmc) has the highest superconducting transition temperature of 20.2 K at ambient pressure. The superconductivity in these technetium-borides mainly originate from the coupling between the dominant presence of Tc-4d electronic states around the Fermi level and the low frequency vibration modes of the technetium-atoms, which is closely analogous to another transition-metal-boride MoB$_2$. Our calculations not only identified superconducting TcB$_2$ (P6/mmm) of the same crystal structure as MgB$_2$ and MoB$_2$, but also discovered series of superconducting technetium-borides with diversified crystal structures. This work proves the rich structures and stoichiometries in superconducting technetium-borides at high pressures, thus sheds lights on the necessity of extended researches in discovery of new superconducting transition-metal borides.

\section{Acknowledgement}

P. Z., Y.D. Q. and S.X. Y. designed the project; X.R. T. conducted the structure searches; P. Z. and A.Q. Y. calculated the electronic structure, the phonon spectra and the superconducting transition temperatures; all authors prepared the manuscript together. This work is supported by the National Natural Science Foundation of China No. 11604255 and the Natural Science Basic Research Program of Shaanxi No. 2021JM-001. Shuxiang Yang is supported by the Key Research Projects of Zhejiang Lab (Grants No. 2021PB0AC02). The computations are performed at the TianHe-2 national supercomputing center in Guangzhou and the HPC platform of Xi’an Jiaotong University.

\bibliography{refs.bib}

\end{document}